\documentclass{article}

\usepackage{arxiv}

\usepackage[utf8]{inputenc} % allow utf-8 input
\usepackage[T1]{fontenc}    % use 8-bit T1 fonts
\usepackage{hyperref}       % hyperlinks
\usepackage{url}            % simple URL typesetting
\usepackage{booktabs}       % professional-quality tables
\usepackage{amsfonts}       % blackboard math symbols
\usepackage{nicefrac}       % compact symbols for 1/2, etc.
\usepackage{microtype}      % microtypography
\usepackage{lipsum}		% Can be removed after putting your text content
\usepackage{graphicx}
\usepackage{natbib}
\usepackage{doi}
\usepackage{listings,multicol}

\title{Storage and Authentication \\ of Audio Footage for IoAuT Devices\\ Using Distributed Ledger Technology}

\author{Srivatsav Chenna\thanks{Both authors contributed equally.} \\
		International Audio Laboratories Erlangen\\
	University of Erlangen-Nuremberg\\
	Erlangen, Germany \\
	\texttt{srivatsav.chenna@fau.de} \\
	\And
	\href{https://orcid.org/0000-0001-6758-4491}{\includegraphics[scale=0.06]{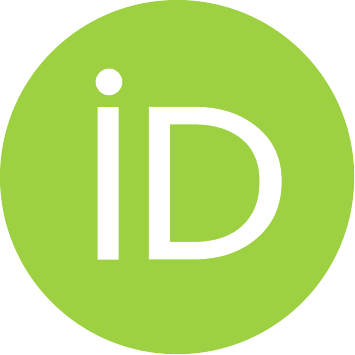}\hspace{1mm}Nils Peters} \\
	International Audio Laboratories Erlangen\\
	University of Erlangen-Nuremberg\\
	Erlangen, Germany\\
	\texttt{nils.peters@fau.de} \\
}

\renewcommand{\shorttitle}{\textit{arXiv} Template}

\hypersetup{
pdftitle={Storage and Authentication of Audio Footage for IoAuT Devices Using Distributed Ledger Technology},
pdfsubject={cs.SD},
pdfauthor={Srivatsav Chenna, Nils Peters},
pdfkeywords={Audio authentication, Audio fingerprinting,Internet of Audio Things, Blockchain},
}

\begin{document}
\maketitle

\begin{abstract}
	Detection of fabricated or manipulated audio content to prevent, e.g., distribution of forgeries in digital media, is crucial, especially in political and reputational contexts. Better tools for protecting the integrity of media creation are desired.  Within the paradigm of the Internet of Audio Things (IoAuT), we discuss the ability of the IoAuT network to verify the authenticity of original audio using distributed ledger technology. By storing audio recordings in combination with associated recording-specific metadata obtained by the IoAuT capturing device, this architecture enables secure distribution of original audio footage, authentication of unknown audio content, and referencing of original audio material in future derivative works. By developing a proof-of-concept system, the feasibility of the proposed architecture is evaluated and discussed. 
\end{abstract}

% keywords can be removed
\keywords{Audio authentication \and Audio fingerprinting \and Internet of Audio Things \and Blockchain}

\section{Introduction}
\begin{figure}
    \centering
    \includegraphics[width=0.6\columnwidth]{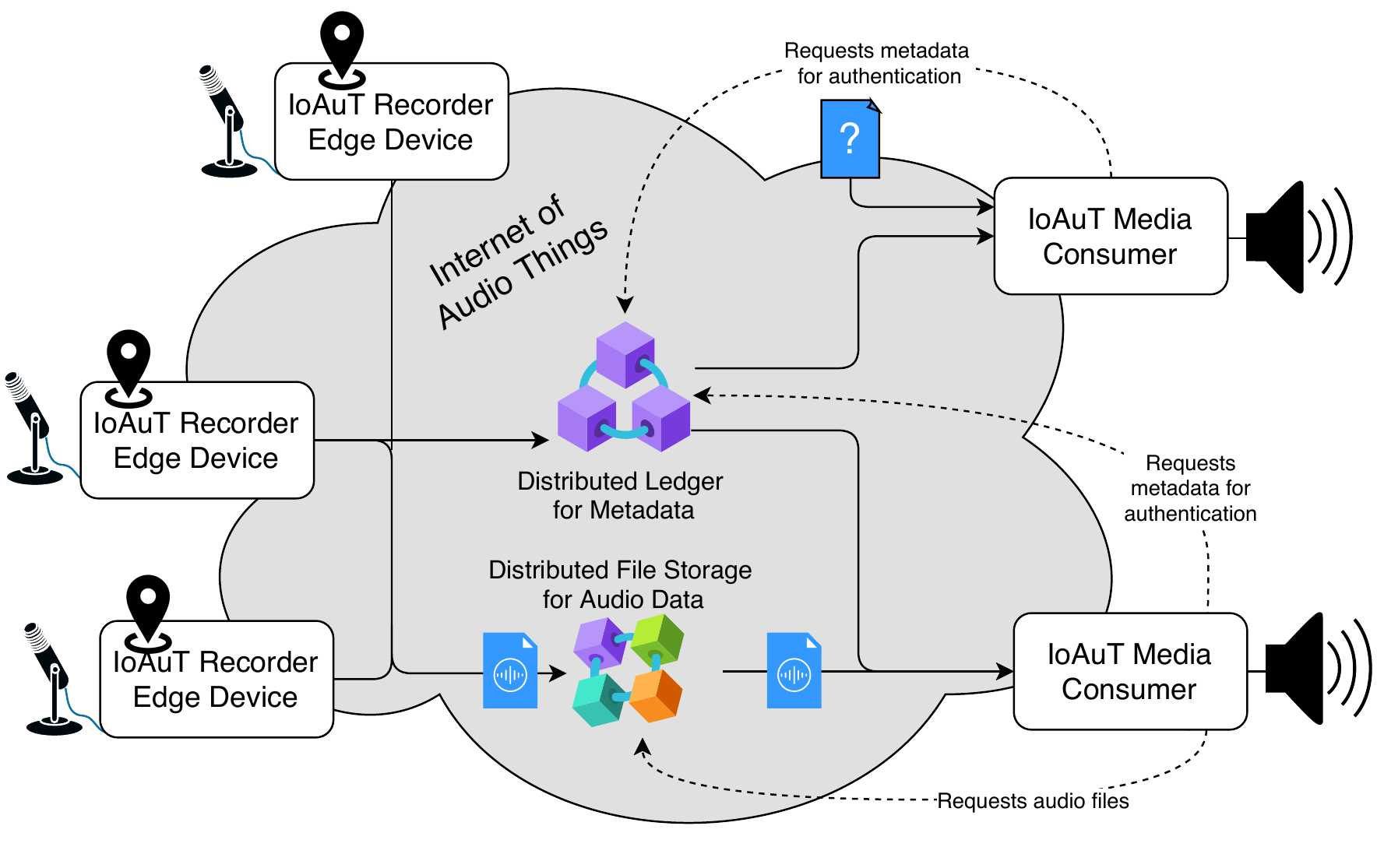}
    \caption{Principle of using distributed ledger technology in combination with IoAuT devices. The IoAuT recorder captures audio and uploads the audio file together with meaningful metadata of the capture (e.g., time, geolocation, acoustic fingerprint) to the distributed ledger. Once the uploaded data are part of the distributed ledger, an IoAuT media consumer can access and use these metadata to authenticate audio footage.} 
    \label{fig:concept}
\end{figure}

The ease of media content distribution via social networks combined with the growing availability of powerful machine-learning tools that are designed to manipulate or synthesize deceptive media footage (a.k.a. deepfakes) created a perfect storm for misrepresentation at scale. To prevent mass distribution of misleading media content, the need for content authentication is greater than ever. \\
Envisioned for audio capturing devices that have internet connectivity and can provide geolocation information (e.g., via GPS), we discuss in this paper the concept to store captured audio and associated metadata in a distributed Internet of Audio Things (IoAuT) network (see Figure \ref{fig:concept}). The advantages due to the distributed and open nature of the network combined with the permanent storage properties are manifold: The architecture enables secure distribution of original audio footage; authentication of unknown audio content; and reference to the original audio material in future derivative audio productions.         

The rest of this paper is organized as follows. In Section \ref{sec:SOA} different technologies for audio content authentication are reviewed. Section \ref{sec:DLT} presents the concept of distributed ledger technology. In Section \ref{sec:Prototype}, we introduce our system architecture. In Section \ref{sec:deployment}, we present a proof-of-concept deployment. The paper concludes with a discussion and outlook.    

\section{Authentication of Media Content}\label{sec:SOA}
Detection of tampered or fabricated audio recordings is a discipline within the domain of audio forensics (see, e.g., \cite{maher_principles_2018}). 
Typical methods of audio tampering or audio forgery include deletion, insertion, substitution, splicing, and copy-move \cite{maksimovic_copy-move_2019} of audio content.

Traditionally, signal processing approaches are used for detecting tampered audio and may require manual review by experts (see \cite{brixen_techniques_2007,nicolalde_evaluating_2009}); but also novel machine-learning approaches have been recently proposed (e.g., \cite{ali2017automatic,wang2020deepsonar}).
Nowadays, with the increased availability of machine-learning based audio synthesis tools, the knowledge and efforts needed to generate authentic sounding audio footage, e.g., to mimic a person's vocal characteristic \cite{adobe2016,respeecher}, decreased significantly. Although methods are being developed to detect deepfake audio signals \cite{10.1007/978-3-030-61702-8_1}, deepfake methods evolve accordingly to undermine detection methods \cite{hussain2021adversarial}. 

\subsection{Watermarking}

Watermarking embeds information (e.g., metadata) directly into the audio signal by utilizing methods similar to perceptual audio. Rather than trying to detect fabricated audio recordings after the audio has been distributed, an alternative approach could be to enable authentication of audio recordings via metadata embedded prior to distribution. For the content authentication, a dedicated decoder can recover those metadata, which may be relevant for content verification.  

As summarized in \cite{xiang2017digital}, common properties of watermarking techniques include robustness, security, perceptibility, capacity, and complexity. Some watermarking techniques are able to extract the signature and restore the original waveform (i.e., reversibility). Depending on the use case, known watermarking techniques trade-off these properties against each other. Primarily, the perceptibility is affected by the size of the embedded data payload (capacity) and by the robustness against unintentional or malicious attacks. For battery-powered (mobile) IoAuT devices, the complexity of the watermarking embedding becomes also crucial.

\subsection{Acoustic Fingerprinting}
A third option considered for authentication is based on acoustic fingerprints. Fingerprints describe the perceptually most relevant acoustic components of an audio file. The fingerprints can be stored in a database together with information about the audio file. To identify an unknown audio file, its fingerprint is computed and matched against that database. Compared to watermarking, fingerprints are less vulnerable to attacks and do not introduce signal alteration/distortions \cite{cano2005audio}. 

In contrast to the hashing values of computer files, the acoustic fingerprint of an audio recording is computed solely on the audio data. Other auxiliary data in the audio file, such as header information and other metadata are not relevant. Also, depending on the algorithm, the acoustic fingerprint can be robust to basic audio processing such as gain changes, but can also withstand moderate audio compression.

\section{Distributed Ledger Technology}\label{sec:DLT}

The use of distributed ledger technologies (DLT) is popularized due to Blockchain and a growing interest in cryptocurrency trading, such as Bitcoin, Ethereum, and numerous other virtual currencies. However, its principles of cryptography, decentralization and consensus enable many other applications \cite{pilkington2016blockchain}. 

Specifically, these applications require and benefit from secure, permanent, and non-reversible information storage. For instance, the authors of \cite{bhowmik_jpeg-blockchain_2018} proposed a blockchain framework for digital image transactions tailored for its use in galleries, museums, and other cultural heritage institutions. Also JPEG experts have considered the support for DLT in their image file format \cite{jpeg_blockchain_2019}. To the authors' knowledge, no audio-specific DLT framework for content authentication has been proposed yet.

\begin{figure}
    \centering
    \includegraphics[width=0.6\columnwidth]{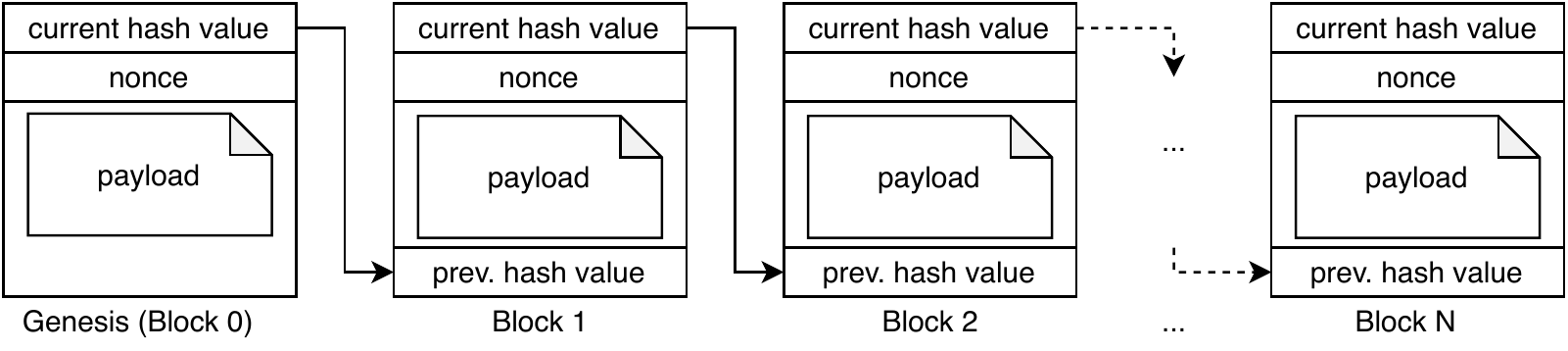}
    \caption{A simplified blockchain}
    \label{fig:blockchain_simple}
\end{figure}

A blockchain system consists of a distributed peer-to-peer network and a transaction ledger which, along with cryptographic hash functions, provide tamper-proof storage of data. It provides a consensus mechanism between the peers to validate the data stored on the ledger. A simple blockchain structure is shown in Figure \ref{fig:blockchain_simple}.

\subsection{Blockchain Overview}A blockchain consists of several parts: the actual payload (a.k.a. the transaction data), the hash of the current block, the hash of the previous block, and the so-called nonce value. The first block of the blockchain is the genesis block, it forms the base for all following blocks. A nonce is an abbreviation for "number only used once", referring to the unique number a blockchain miner needs to discover before solving for a block in the blockchain. Nonce is difficult to find as it requires high computational power to solve. A cryptographic hash is generated with the nonce for the block that serves as a digital signature for the block. When the transaction data in a block is manipulated, the cryptographic hash of the block changes and would consequently differ from the stored hash. Furthermore, storing the cryptographic hash of every block in its following block prevents replacing existing chain elements with forged ones. As soon as the blockchain is distributed across the network, the stored transaction data are inherently secured because a single user cannot change the existing blockchain.

\subsection{Transaction}A transaction is the process of storing the payload data permanently in the blockchain. When a transaction is added to the blockchain, it is verified and distributed to each node in the blockchain network. Table \ref{tab:blockchainStrcture} shows the transaction data stored in our proposed blockchain application. 

\subsection{Mining}Mining is the process of adding new blocks to the blockchain. Miners solve a "Proof of Work", which is a complex math problem to find the nonce value of the block. Only after finding the nonce, the block is added to the block chain. Any manipulation to the block earlier in the chain requires the miner to re-mine not only the manipulated block but also all the blocks following it. Because of the complexity in mining, manipulating data on the blockchain is not considered feasible \cite{10.1007/978-981-13-6621-5_5}. After a block is mined, it is validated by all the nodes in the blockchain network and then added to the blockchain.

\subsection{Consensus}
Consensus is a decision-making mechanism associated with approving transactions in a blockchain. In the current prototype, Proof of Work is used for consensus. A block proposed by a miner is accepted into the blockchain when all the transactions are validated by more than the majority of the nodes on the network. To validate a block, the mining node on the network has to check if the miner proposing the block has solved Proof of Work and if all the hashes linking the blocks are accurate. %Additionally, in the proposed system prototype the mining nodes verify the signature, duration and number of channels channel of the audio footage.        \red{Finish me}        

\subsection{Adding a New Block}
To add a new block to the chain, a miner has to perform the following steps:
\begin{enumerate}
\item Search in the available chains in the blockchain network and adopt the 'best' chain (best chain is usually defined by the longest available valid chain)
\item Check if hash of each block accurately points to the previous block (i.e., verify chain)
\item Search the blockchain network for novel transactions that are not included in the chain yet
\item Create a new block with the new transactions and include the hash of the last block of the chain in the new block 
\item Solve Proof-of-Work algorithm and propose new block
\item Nodes in the blockchain network validate the proposed block
\item New block is added to the chain and announced to the network
\end{enumerate}

%\red{explain some blockchain fundamentals: Block, Transaction, Mining, Proof-of-work, proposing new blocks?}

%\red{How to create a New Block?
%1. Verify if the transactions are valid
%3. Select the hash value of the most recent block and insert it into the new block as a link
%4. Solve the proof of work algorithm
%5. Add the new block to the local block chain and propagate it to the network.}

\section{System Prototype} \label{sec:Prototype}
To study the feasibility of using DLT for content authentication in the IoAuT context, we developed a prototype based on two distributed data technologies: a blockchain and a distributed file storage system (see Figure \ref{fig:concept}). 
The  distributed file storage system is used to store the audio recordings. We use the Interplanetary File System (IPFS) for this purpose because contrary to the blockchain, IPFS allows storing of large files (i.e. audio data). To access the stored audio data, one needs to know the corresponding cryptographic hash value (the IPFS-hash).
The blockchain stores associated metadata, an acoustic fingerprint, as well as the hash information to access the audio recordings from the IPFS. The implementation was done in Python and Flask and is based on the blockchain framework \cite{pythonBs}.

A Python-integrated IPFS framework was added to establish connection between the blockchain servers and the IPFS servers. Necessary audio handling algorithms to obtain audio device and user information (MAC address, GPS info, device maker and model) and to compute acoustic fingerprints are implemented in Python. The Flask framework is used to run the blockchain servers. Flask enables communication between the nodes using HTTP/HTTPS. An application programming interface (API) and a command-line interface is implemented to interact with the blockchain servers (create transactions, mine blocks, etc).

We posit that at the moment of capturing audio, available information about the recording time, recording place, and recording device is prime metadata for content authentication. Immediately after capturing audio (e.g., recording an interview), the audio file and its metadata is registered with the blockchain and uploaded to the distributed file system.

%\subsection{Distributed Audio Storage}

\subsection{The Blockchain Payload}
A transaction into the blockchain consists of a number of relevant metadata describing the associated audio file and its origin (see Table \ref{tab:blockchainStrcture}). This payload also includes the unique content ID that links the blockchain to the actual audio file stored within the IPFS. 

\begin{table}[!h]
\caption{Metadata within the blockchain payload}
    \centering
    \begin{tabular}{l|l}
   \hline \hline
      Data field & Description \\ \hline 
      version & Version number of the payload\\
      recFileName & Name of the audio file\\
        recTimestamp & Time of the recording\\
        recDuration & Duration of the audio recording\\
        recNumChannels & Number of audio channels\\
        deviceMaker & Recording device manufacturer\\
        deviceModel & Recording device model\\
        deviceMacAdd & MAC Address of IoAuT recorder \\
        deviceGpsInfo & GPS geolocation data \\
%        author & Name of creator (optional) \\
        ipfsHash  &  Link to the audio file in the IPFS\\
        contentId & Identifier as part of the blockchain\\
        recSignature & Acoustic fingerprint of the audio file\\
        \hline \hline
    \end{tabular}
    
    \label{tab:blockchainStrcture}
\end{table}

\textbf{version:} A version number for the blockchain payload. This enables future improvements and additions while maintaining backwards compatibility.

\textbf{recFileName:} Name of the audio file. At the moment the blockchain supports .wav, .mp3, and .m4a file types.

\textbf{recTimestamp:} The date and time the recording finished.

\textbf{recDuration:} The length of the recording in seconds.

\textbf{recNumChannels:} The number of recorded audio channels; must be one or more.

\textbf{deviceMaker:} The name of the manufacturer of the recording device.
 
\textbf{deviceModel:} The model name of the audio recorder. 

%\textbf{deviceSerialNumber:} Serial number of the recording device. \red{is this a good idea?}

\textbf{deviceMacAdd:} The Media Access Control (MAC) Address of the IoAuT recording device. The MAC Address is a unique identifier of the network hardware of the IoAuT device, e.g., 00:1A:44:11:3B:B7. 

\textbf{deviceGpsInfo:} If the IAoT recording device is equipped with GPS capabilities,the geolocation information at the time the recording was finished can be stored in the blockchain (latitude and longitude information in decimal degrees).

\textbf{contentId:} A unique identifier according to \cite{leach2005universally} to identify the audio file within the distributed file system. The contentId is also embedded as ID3v2 metadata into the audio file. ID3v2 are supported by various audio file formats, including .wav, .mp3, or .m4a. Even when the audio file is renamed and distributed elsewhere, the contentId is preserved and links to its unique blockchain record. 

\textbf{ipfsHash:} The cryptographic hash of the distributed file system IPFS that links to the storage location of the audio file.

\textbf{recSignature:} The acoustic fingerprint of the audio data. Instead of a file hash value, we deliberately opted to use an acoustic fingerprint signature for authentication. Compared to computing a hash value over the entire file, acoustic fingerprints are computed in the audio domain. 

Once the blockchain is mined, a block will also contain the hash value to the previous block, its own hash value, the nonce, and a timestamp of the moment the the block was mined.
An example of our blockchain can be seen in Listing \ref{lst:chain2}.

%\subsection{Authentication Process}
%\red{TODO: Describe how the authentication process works and why this is trustful? \url{https://oxygene.sk/2011/01/how-does-chromaprint-work}}

\section{Proof-of-Concept Deployment}\label{sec:deployment}
To test the distributed authentication concept, we deployed a DLT system with five blockchain nodes. 

\begin{figure}
    \centering
    \includegraphics[width=0.5\columnwidth]{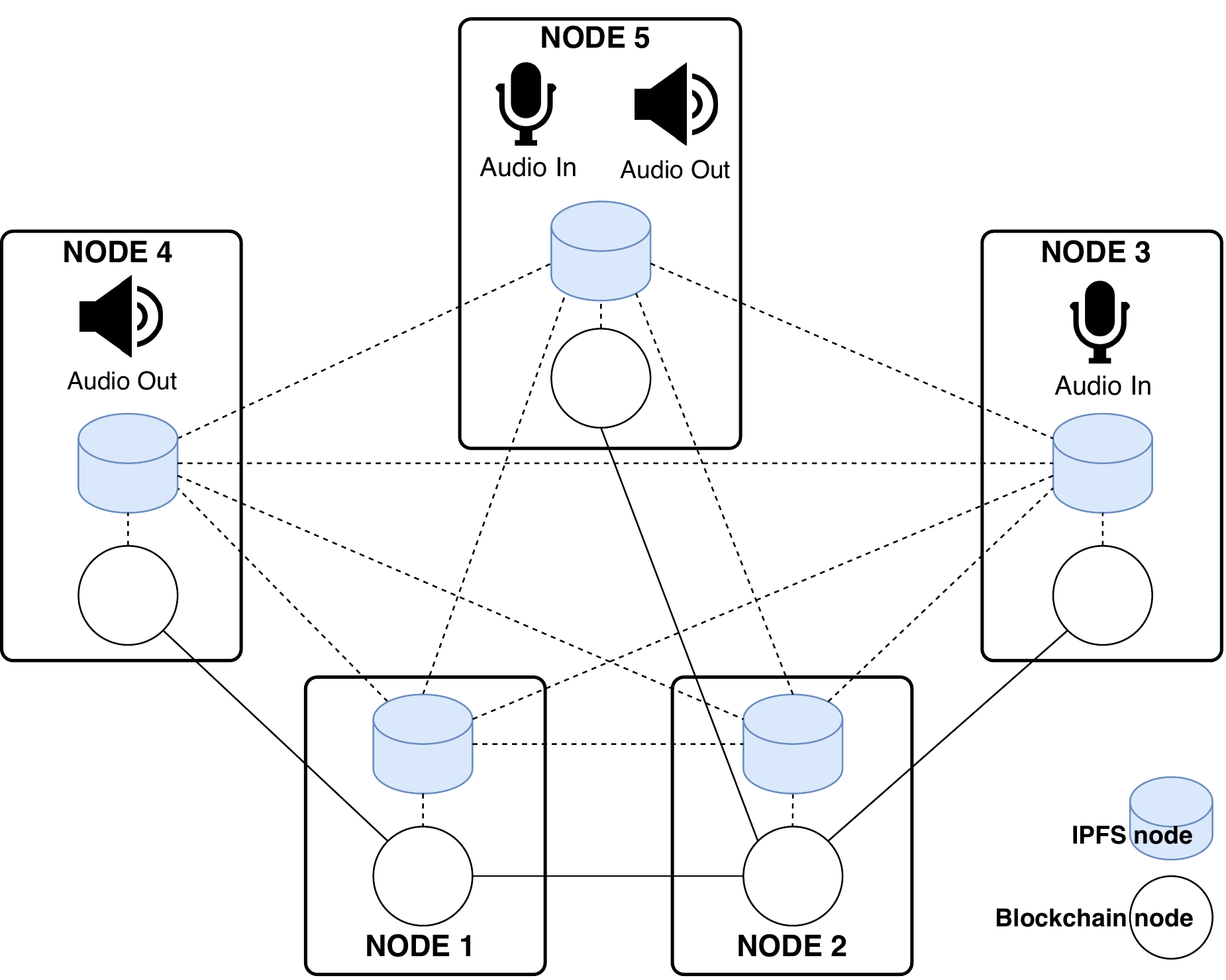}
    \caption{Schematics of the Proof-of-Concept deployment}
    \label{fig:overview}
\end{figure}

An overview of the proof-of-concept network is depicted in Figure~\ref{fig:overview}. The network is extendable by registering additional nodes. 
Each of the devices is internally connected to the blockchain and the IPFS network.
Each blockchain node is able to propose new blocks, send new transactions, and have complete history of the blockchain. The blockchain nodes are connected to each other in a peer-to-peer network and are able to exchange updates with respect to the changes in the blockchain.   

%Each blockchain node is internally connected to an IPFS node. %A private IPFS network was configured so that only the nodes in the private network can share audio data. 

This proof-of-concept deployment serves as a small-size example to approximate an real-world IoAuT scenario which consists of a plurality of devices with different audio processing capabilities:

\textbf{Node 1 and 2} act as generic (server) nodes in the distributed network and are implemented on desktop computers. This type of node does not have any audio periphery but are important to verify and mine the available transactions and to add them to the blockchain. 

\textbf{Node 3} serves as a prototype of an IoAuT-enabled recorder. It is implemented on a single board low-power ARM computer and equipped with a stereo microphone. Additional sensors help to determine its geolocation. Recorded audio is compressed using Advanced Audio Coding (AAC) into the MP4 file format. The MP4 file is uploaded to the IPFS network as soon as the recording is fully captured and compressed, and the metadata payload specified in Table \ref{tab:blockchainStrcture} is registered as a novel transaction with the blockchain.  

\textbf{Node 4} is a prototype of an IoAuT-enabled media player. It is implemented on a single board low-power ARM computer and is connected to a loudspeaker. It is programmed to: a) monitor the blockchain and to download the audio footage referenced in the most recent block; b) verify its authenticity via comparing its fingerprint; and c) playback the verified audio file. 

\textbf{Node 5} is an example of a mobile multi-purpose device. It has the combined audio capabilities of node 3 and node 4. It can record and feed new audio material into the blockchain, and can also consume and authenticate media from the blockchain.

\subsection{Interactions with the Network}
This section describes the functionalities for how to interact with the distributed network.
 \subsubsection{Contributing Audio Footage}
The devices that have audio capturing functionalities (node 3 and node 5) can upload the captured footage as an AAC-encoded audio file to the IPFS node. The blockchain node generates a unique 32 hexadecimal digit identifier (see contentId in Table \ref{tab:blockchainStrcture}) and embeds it as the audio copyright ID3v2 metadata into the AAC file. Audio encoding and ID3v2 metadata writing is achieved using the FFmpeg tool \cite{FFmpeg}. The fingerprint of the audio file is generated using the acoustic fingerprinting algorithm \cite{acoustid}. Other Python routines determine the device information (deviceMacAdd, deviceGeoInfo, deviceMaker, deviceModel) and metadata about the audio footage (recTimestamp, recDuration, recNumChannels). The blockchain node creates a transaction with the payload described in Table \ref{tab:blockchainStrcture}. In this proof-of-concept architecture, a block contains one single transaction. Bundling of multiple transactions into one block would be possible, but is not considered in this study.

%Solved %\red{Question: is the audio uploaded to the blockchain node, or the IPFS node? We need to be consistent and precise here} %: \red{its uploaded to the IPFS node, I fixed the mistake}
%solved: \red{TODO: Size if Acoustic Signature depends on audio file length? what did we test?}

\subsubsection{Mining Transaction} \label{sec:mining}
Although any node is able to mine transactions, the mining operation is sought to be primarily executed by the servers (node 1 and node 2). When a node mines a transaction, all the transaction data is uploaded into the blockchain after verification (proof is valid and the hashes linking the chain are valid). 
To prevent fake or corrupted metadata to be uploaded on the blockchain, the miners also verify the acoustic fingerprint, duration, and the number of channels of the audio footage. After successful mining, the updated chain is distributed to all the nodes in the network.

Listing \ref{lst:chain2} shows the blockchain as a Python JSON object. 
The first block shows the blockchain genesis block, which contains no transaction data. Its previous\_hash value is zero.
The genesis block is followed by two blocks with recorded transactions. Each transaction contains the metadata payload as described in Table \ref{tab:blockchainStrcture}). Listing \ref{lst:chain2} also shows the blockchain-specific nonce value and the linkage between two consecutive blocks via the hash and previous\_hash values.

% "length": 1, 
%\begin{lstlisting}[frame=single,basicstyle=\footnotesize,breaklines,caption={Genesis block. Hash value is shortened for readability.  },captionpos=b,label={IoAuT}]
%"chain": [
%{
% "index": 0, 
% "transactions": [], 
% "timestamp": 0, 
% "": "0", 
% "nonce": 0, 
% "hash": "6db...0c9"
%}]
%\end{lstlisting}

%, "peers": [
% "http://131.188.158.161:8000/", 
% "http://131.188.158.164:8000/", 
% "http://131.188.158.91:8088/"]

% "length": 2, 
%[numbers=left,xleftmargin=3em, ] , 
% 
\begin{lstlisting}[float=tp,frame=single,basicstyle=\footnotesize,breaklines,captionpos=b,label={lst:chain2},caption={Blockchain after two valid transactions. Some values were shortened for readability.}] 
"chain": [
{
 "index": 0, 
 "transactions": [], 
 "timestamp": 0, 
 "previous_hash": "0", 
 "nonce": 0, 
 "hash": "6db...0c9"
}, 
{
 "index": 1, 
 "transactions": [{
    "version": "1", 
    "recFileName": "interview_J_Doe.m4a", 
    "recDuration": 120.0,  
    "recTimestamp": 1616499571.0820450,
    "recNumChannels": 2
    "deviceMaker":"Raspberry Pi Foundation"
    "deviceModel": "Raspberry Pi 4 Model B"
    "deviceMacAdd": "00:05:9a:3c:7a:00", 
    "deviceGpsInfo": [49.591, 11.0078], 
    "ipfsHash": "Qmb...4JT",
    "contentId": "9d7...c69f", 
    "recSignature": "b'A...AKA'"}], 
 "timestamp": 1616499594.3911936, 
 "previous_hash": "6db...0c9", 
 "nonce": 149, 
 "hash": "001...9be"
},
{
 "index": 2, 
 "transactions": [{
    "version": "1", 
    "recFileName": "highway_traffic.m4a", 
    "recDuration": 61.0,  
    "recTimestamp": 1616499589.5719413, 
    "recNumChannels": 1
    "deviceMaker": "Google, Inc."
    "deviceModel": "Pixel 2"
    "deviceMacAdd": "8A:F3:2A:D3:10:72", 
    "deviceGpsInfo": [49.5444, 11.0177] 
    "ipfsHash": "2fn8...QsN",
    "contentId": "417...59b", 
    "recSignature": "AQA...SQQ"}], 
 "timestamp": 1616499604.6480861, 
 "previous_hash": "001...9be", 
 "nonce": 74, 
 "hash": "b07...cb8"
}]
\end{lstlisting}

%
%, 
%"peers": [
% "http://131.188.158.161:8000/", 
% "http://131.188.158.164:8000/", 
% "http://131.188.158.91:8088/"]
%
\subsubsection{Audio Consumption}
%\red{Question: does the node actively need to request the valid blockchain first?}
% \red{The validation only occurs during mining and adding a new block }

Media consumption devices (i.e., node 4 and node 5) are able to retrieve an audio file via its contentId from the network.
Once the audio file has been downloaded, the device can calculate the audio fingerprint and compare it to the associated fingerprint registered in the blockchain. If the two fingerprints differ, the user is notified that the audio is not genuine. In this verification process, additional information of the metadata payload can also be verified, e.g., duration and number of audio channels.

\subsubsection{Authentication of Unknown Files}

When the media playback device is faced with an unknown audio file from another source, rather than downloaded from the IPFS, the media playback node can search the blockchain to find a match with the fingerprint and the file duration of the unknown file. A successful match would authenticate the unknown file, thus revealing the additional device-specific metadata from the blockchain payload.

Alternatively, the matching transaction payload can also be identified if the unknown file has stored the contentId within its ID3v2 copyright metadata. The media playback device could then identify the matching blockchain transaction and verify the audio metadata accordingly.

\subsection{Metadata Robustness} \label{sec:robustness}
The audio fingerprinting algorithm was tested for its robustness with respect to audio manipulation.
The audio was manipulated using the SoX audio tool \cite{sox}. Fingerprints of the original and manipulated audio material are computed as in our proof-of-concept implementation using fpcalc \cite{acoustid}. We applied the following audio manipulations in a variety of configurations: trimming (i.e. shortening the audio file), gain change, time shift, and pitch shift. A speech recording was used as the original content. The results of these experiments show that the fingerprint signature changed in all tested conditions (see Table \ref{tab:robustness}). These results suggest robust forge detection of the audio footage in the proposed proof-of-concept system.

\begin{table}[!h]
\caption{Modifications and resulting change in the audio signature}
    \centering
    \begin{tabular}{lrc}
   \hline \hline
Method & Modification & Signature \\
& Strength & Change \\
\hline 
 Trim     & 0.1 s       & Yes   \\
          & 1.0 s       & Yes   \\
          & 3.0 s      & Yes   \\
          & 10.0 s       & Yes   \\
 \hline
Amplification & ±1 dB & Yes \\
          & ±3 dB       & Yes \\
          & ±10 dB       & Yes  \\
\hline          
Time shift & ±1 \%     & Yes \\
           & ±10 \%      & Yes\\
           & ±50 \%     & Yes \\
\hline 
Pitch shift & 1 cent & Yes \\
 & 10 cents & Yes \\
 & 100 cents & Yes \\
\hline\hline
    \end{tabular}
    
    \label{tab:robustness}
\end{table}

%TODO: Scoring nodes      
\subsection{Further Security Observations}
%Integrity relates to the reliability and trusthworthiness of the data.
% privacy protection 
With respect to blockchains, security experts primarily focus on integrity, scalability, and privacy protection \cite{10.1007/978-981-13-6621-5_5}. In the scope of this specific proof-of-concept deployment, we have already observed the following security aspects in the proof-of-concept deployment.

Data Integrity relates to the reliability and trusthworthiness of the data, and computing the cryptographic hash signature value of each block is the basis for blockchain data integrity.
Our proof-of-concept implementation utilizes the SHA-256 secure hash algorithm which efficiently creates a unique 256-bit hash value from the blockchain data. The SHA-256 algorithm provides suitable trade-offs in terms of computational effort (especially for mobile IAoT devices), storage requirement within the block (256 Bits), and the probability for a hash value collision which is about $4.3\cdot10^{-60}$ for a chain comprised of one billion blocks. For the same advantages, the ipfsHash that connects to the audio data is  generated with SHA-256 too.

Each block has a fixed maximum size, e.g., 1 MB. To guarantee that the complete set of recording metadata is stored, the size of the transaction data must not exceed the maximum blocksize. The largest metadata usually originates from the acoustic fingerprint. In the current implementation, the size of the acoustic fingerprint scales with the recording duration and the number of audio channels. Due to the built-in data compression (see \cite{acoustid}), the fingerprint size is efficiently reduced, but somewhat content dependent. In our tests, a one-hour radio interview resulted in a fingerprint of about 80 kB, whereas another recording of the same length yielded a fingerprint of about 45 kB. These measurements demonstrate that the blocksize does not impose a practical limitation for most recording scenarios. 

Privacy protection is an increasing concern in the connected world. For our initial proof-of-concept study, all recording metadata are stored in plaintext. This allows other nodes to use the recording metadata to verify an uploaded recording (see more in Section \ref{sec:ChainOfTrust}), but also enables others to obtain this information. Unconditional access to all stored metadata may not be desired with respect to data privacy \cite{wang_survey_2020}. In an improved deployment, possibilities to encrypt certain metadata values (e.g., the recFileName value) will increase data privacy.

%- Are block payloads encrypted? How are the keys managed and revoked?\\
%- logic for resolving blockchain block collisions?\\
%- public or private network?\\
%- IPFS file security?\\
%
%scholars mainly focus on \textbf{integrity}, \textbf{privacy protection}, and \textbf{scalability}.\\
%existence of proof-of-work mechanism and the large number of honest miners make blockchain integrity protected.\\
%cryptographie is the cornerstone of the blockchain technology. Once the hash function or or encryption algorithm is no longer secure, the security of te blockchain will no longer exist. The hash function SHA256 and the encryption algorithm elliptic curve cryptography used fir the blockchain are still safe, but new technologies (quantum computing) its security remains to be discussed. 

\section{Discussion and Future Work}
%\red{TODO}
%
%\subsection{Other Use Cases}
%audio capturing for body cameras?, smart city monitoring?\\
%How to trace evolution of audio content within the BS, e.g, for content trimming, remixing \\

\subsection{Evolution of the Transaction Payload}
The initial design of the blockchain payload shown in Table \ref{tab:blockchainStrcture} is not necessarily in its final state.
Using the version field, evolution of the payload is envisioned.
There may be a desire for additional data fields to store more information worth preserving in a blockchain, e.g., the name of the device user or the serial number of the capturing device. Also, support for time-varying data (e.g., the changing geolocation of a moving recording device) may hold relevant information.
Besides the acoustic fingerprint, the blockchain does not hold acoustic data. For some applications, it may be useful to store some high-level audio descriptors, such as the signal envelope or a transcription via an automatic speech recognition algorithm. Alternatively, a representative short excerpt from the original audio footage (a so-called acoustic thumbnail \cite{gravier2014audio}) could be included.
The computation of the acoustic fingerprint may also change in future payload versions. Our experiments with the current method, however, suggest effectiveness of the fingerprint (see Section \ref{sec:robustness}). Selecting the best acoustic fingerprint processing is outside the scope of this study.

\subsection{Building the Chain of Trust} \label{sec:ChainOfTrust}
The ipfsHash and the acoustic fingerprint stored in the blockchain transaction links the audio data in the IPFS to the trusted metadata in the blockchain. Metadata are relevant for authenticating the audio recording with respect to time, location, and recording device. An obvious concern is trustworthiness of the data. Can the actual audio footage or the metadata being manipulated during upload, e.g., via a man-in-the-middle attack? How might deliberately forged metadata be uploaded into the blockchain? How could audio data be manipulated or deliberately forged at any point in this process?
%PART ONE: DEMONSTRATE HOW YOU HAVE THOUGHT THROUGH THE ACTUAL TECHNICAL PROCESS OF UPLOADING %This would possibly undermine the usefulness of the concept.

A man-in-the-middle attack during the process of contributing data to the storage system is preventable by using the encrypted web protocol HTTPS. 

Potential forms of blockchain attacks include bribe attack, long-rang attack, precomputing attack, Sybil attack and the 51\%-attack. In a bribe attack, the attacker attempts to bribe users in the blockchain network to add the transactions proposed by the attacker. 
A long-range attack happens when attackers build an alternative blockchain starting from the origin block. A precomputing attack is carried out when attackers secretly create a chain of new blocks and release them all at once to override the correct blockchain.
In a 51\% attack the majority of the network is taken over in order to erase earlier transactions in the blockchain network.
Because of the computational complexity of the Proof-of-Work consensus, the aforementioned attacks are considered either as impossible or impractical. However, the Proof-of-Work consensus is vulnerable against Sybil attacks in which attackers attempt to take over the network by creating malicious nodes to out-vote the honest nodes. 
However, the cost of this attack is high and grows with the size of the network. Thus, the reward may not compensate the cost.
In summary, the Proof-of-Work consensus mechanism outperforms other consensus mechanisms in terms of security at the cost of computational complexity \cite{10.1007/978-981-13-6621-5_5}.

Importantly, in our application scenario the chain of trust does not start with the blockchain, but with the IoAuT recording devices themselves. 
It is critical to trust the audio signals and metadata (e.g., MAC address) in the transaction payload.
For instance, it was demonstrated that MAC address can be spoofed or that geolocation data can be made inaccurate. %Consequently, the entire ecosystem that creates audio and metadata for the Internet of Sounds needs to be studied and made transparent and verifiable.

To avoid registration into the blockchain of purposely fabricated audio files, devices should only permit blockchain authentication of physical (microphone) inputs rather than audio files from an unknown origin. 
To extend the chain of trust onto the hardware layer, the firmware of the recording device would have to: a) secure the channel from the physical input to the device memory; b) protect the data against external write access; and c) allow a protected data upload. 

%This could be realized when the firmware of the recording device creates a protected media storage that can is to the data upload.
%Also, by first checking for the existence of a previously registered acoustic fingerprint in the blockchain, the upload process could prevent attempts of adding redundant (and likely false or duplicate) audio files to the system.
To prevent forged metadata to be included in the blockchain, tampering detection methods may be applied as part of the mining operation. The mining nodes in our proof-of-concept network (nodes 1 and 2 in Figure \ref{fig:overview}) can process audio and metadata payload to detect forgery or tampering. If a considerable number of the processing nodes (e.g., the majority of all miners) have verified the data, the transaction has permission to be registered with the blockchain. In our proof-of-concept network, we have integrated a low-complexity data verification process into the mining operation of node 1 and node 2. Here the metadata are checked for data errors and plausibility.
More sophisticated detection methods, such as a machine learning model for fake voice detection \cite{10.1007/978-3-030-61702-8_1}, could augment this verification step in the future.

\subsection{Future Work}
While experimenting with the proof-of-concept deployment, we observed that all audio files are accessible for all users of the IPFS network. This accessibility may not be feasible or desired. Therefore, future work could address data distribution rules and access restrictions. Technical proposals for such modification exist \cite{8726493}. 

To widen the initial scope of storing and authenticating audio footage, it would be interesting to explore how the proposed architecture can be expanded to track how and where specific audio footage is being used. What happens with the audio files stored in our system of IPFS network and blockchain? How are they being used or altered, e.g., to produce media sound bites, other sounds or musical pieces? Where are they being used and why? Besides the initial application for audio authentication, this may even help in other use cases, e.g.,  musicians and musicologists to better trace the origin of musical fragments by using the Internet of Audio Things. 

\section{Summary and Conclusion}

We presented a storage and authentication framework for audio footage in an IoAuT context. This framework is based on a combination of blockchain for secure and permanent metadata registration and the IPFS peer-to-peer network for a distributed audio data storage. To the authors knowledge, this is the first audio-specific distributed storage framework specified for authentication needs.
We proposed a system architecture with an extendable metadata structure and experimented with this architecture in a proof-of-concept deployment. The proposed architecture can be an important part for enabling a trustworthy authentication system for recorded audio material. For a reliable end-to-end system, the audio data as well as the metadata that are registered with the blockchain must be trustworthy and secured from malicious attacks. Therefore the blockchain mining operation includes an audio and metadata verification process.

\bibliographystyle{unsrtnat}
\bibliography{references}  %%% Uncomment this line and comment out the ``thebibliography'' section below to use the external .bib file (using bibtex) .





% \end{thebibliography}

\end{document}